\documentclass{amsart}[10pt]
\usepackage{amsmath, amsfonts, amsthm, amssymb}
\usepackage{subfigure}
\usepackage{stmaryrd}
\usepackage[dvips]{graphicx}
\font\twelvemsb=msbm10 at 12pt
\newfam\msbfam
\textfont\msbfam=\twelvemsb

\numberwithin{equation}{section}
\theoremstyle{definition}

\newcommand{\D}{\mathrm{d}}
\newcommand{\E}{\mathrm{e}}

\begin{document}
\title{Variance dispersion and correlation swaps}
\author{Antoine Jacquier}
\address{Department of Mathematics, Imperial College London and Zeliade Systems, Paris}
\email{ajacquie@imperial.ac.uk.}
\author{Saad Slaoui}
\address{Citigroup}
\date{}

\begin{abstract}
In the recent years, banks have sold structured products such as worst-of options, Everest and Himalayas, resulting in a short correlation exposure. 
They have hence become interested in offsetting part of this exposure, namely buying back correlation. 
Two ways have been proposed for such a strategy : either pure correlation swaps or dispersion trades, 
taking position in an index option and the opposite position in the components options. 
These dispersion trades have been set up using calls, puts, straddles, variance swaps as well as third generation volatility products.
When considering a dispersion trade using variance swaps, one immediately sees that it gives a correlation exposure. 
Empirical analysis have showed that this implied correlation was not equal to the strike of a correlation swap with the same maturity. 
The purpose of this paper is to theoretically explain such a spread. 
In fact, we prove that the P\&L of a dispersion trade is equal to the sum of the spread between implied and realised correlation - 
multiplied by an average variance of the components - and a volatility part. 
Furthermore, this volatility part is of second order, and, more precisely, is of volga order. 
Thus the observed correlation spread can be totally explained by the volga of the dispersion trade. 
This result is to be reviewed when considering different weighting schemes for the dispersion trade.
\end{abstract}
\maketitle

\section{Introduction}
For some years now, volatility has become a traded asset, with great liquidity, both on equity and index markets. 
Its growth has been such that some options on it have been created and traded in huge quantities. 
Indeed, variance swaps are very liquid nowadays for many stocks, and options on variance and on volatility have been the subject of several research, 
both from academics and practitioners. 
Furthermore, these products have given birth to positions on correlation, which had to be hedged. 
Hence, products such as correlation swaps have been proposed to answer these needs. 
Our purpose here is to compare the fair correlation priced in a correlation swap and the implied correlation of a dispersion trade. 
Indeed, a dispersion trade can be built upon variance or gamma swaps, hence creating an almost pure exposition to correlation, 
independent of the level of the stock. 
Furthermore, we will explain the observed spread between these two correlations in terms of the second-order derivatives of such dispersion trades. We indeed believe that the moves in volatility, both of the index and of its components, have a real impact on the implied correlation.

\section{Constituents of dispersion trades}
In this section, we briefly review the different financial products which will serve as a basis in the construction of dispersion trades. 
We consider a probability space $\left(\Omega, \mathcal{F},\mathbb{P}\right)$ and a stock price process $S:=\left(S_t\right)_{t\geq 0}$ defined on it
such that $\D S_t=\sigma_t \D W_t$ for any $t\geq 0$, with $S_0>0$, and where $\left(W_t\right)_{t\geq 0}$ denotes a standard Brownian motion. 
We assume that the volatility process $\left(\sigma_t\right)_{t\geq 0}$ is smooth enough and not exploding so that a unique solution $S$ exists.
We will consider European options written on $S$, maturity $T$ and risk-free interest rate $r$.
$\mathcal{N}$ will denote the cumulative standard Gaussian probability function and $\phi$ its density. 
For a European option $V$, we will denote $\Psi:=\partial^{2}_{S\sigma} V$ the Vanna, $\Upsilon_{\sigma}:=\partial_{\sigma} V$ the vega, 
$\Upsilon_{v}:=\partial_{\sigma^2}V$ the vega with respect to the variance, $\Omega:=\partial_r V$ the derivative of $V$ with respect tothe interest rate, 
and $\Lambda_{\sigma}:=\partial_{\sigma\sigma}^2 V$ or $\Lambda_{v}:=\partial^2_{\sigma^2\sigma^2}V$. 
We recall the Black-Scholes price of a call option $C\left(S_{0},K,T\right)$ at time $t\in\left(0,T\right)$ on $S$ with strike $K>0$ and maturity $T>0$ (assuming that $\sigma_t$ is constant for all $t\geq 0$):
$$C\left(S_{0},K,T\right)=S_{t}\mathcal{N}\left(d_{1}\right)-K\E^{-r\tau}\mathcal{N}\left(d_{2}\right),$$
with
$$
d_{1}:=\frac{\log\left(S_{t}/K\right)+\left(r+\sigma^{2}/2\right)\tau}{\sigma\sqrt{\tau}},\quad
d_{2}:=d_{1}-\sigma\sqrt{\tau},\quad\text{and}\quad \tau:=T-t.
$$

We will also consider straddle options, which consist of a long call and a long put options with the same characteristics (strike, stock, maturity). A variance swap with a notional of $N$ is a swap on the realised variance of a stock price, so that its payoff is worth
$$\mathbb{V}=N\left(T^{-1}\int_{0}^{T}\sigma^{2}_{t}dt-K_{\mathbb{V}}\right),$$
where $K_{\mathbb{V}}$ is a fixed amount specified in the contract. Referring to~\cite{GS}, the fair price $K_{\mathbb{V}}^{0,T}$ of the variance swap is equal to
$$K_{\mathbb{V}}^{0,T} = 2r
 - \frac{2}{T}\left(\left(\frac{S_{0}\E^{rT}}{S_{*}}-1\right)+\log\left(\frac{S_{*}}{S_{0}}\right)\right)
 + \frac{2\E^{rT}}{T}\left(\int_{0}^{S_{*}}\frac{\D K}{K^{2}}P\left(S_{0},K,T\right)+\int_{S_{*}}^{+\infty}\frac{\D K}{K^{2}}C\left(S_{0},K,T\right)\right),
$$
where $P\left(S_{0},K,T\right)$ represents a European put option and $S_{*}>0$ a liquidity threshold.
Hence, the variance swap is fully replicable by an infinite number of European calls and puts.
Moreover, if we take $S_{*}:=S_{0}\E^{rT}$, such that the liquidity threshold is equal to the forward value of the stock price, the above formula simplifies to
$$K_{\mathbb{V}}^{0,T}=\frac{2\E^{rT}}{T}\left(\int_{0}^{S_{*}}\frac{\D K}{K^{2}}P\left(S_{0},K,T\right)+\int_{S_{*}}^{+\infty}\frac{\D K}{K^{2}}C\left(S_{0},K,T\right)\right).$$
A variance swap is interesting in terms of both trading and risk management as  : \\
\indent{- it provides a one-direction position on the volatility / variance.}\\
\indent{- it allows one to speculate on the difference between the realised and the implied volatility. Hence, if one expects a rise in volatility, then he should go long a variance swap, and vice-versa.}\\
\indent{- As the correlation between the stock price and its volatility has proven to be negative, variance swaps are also a means to hedge equity positions.}\\
\\
From a mark-to-market point of view, the value at time $t\in\left[0,T\right]$ of the variance swap strike maturing at time $T$ is
\begin{align*}
\Pi_{t}^{T} & = \E^{-r\tau}\mathbb{E}_{t}\left(\frac{1}{T}\int_{0}^{T}\sigma^{2}_{u}\D u-K_{\mathbb{V}}^{0,T}\right)\\
 & = \E^{-r\tau}\left(\frac{1}{t}\frac{t}{T}\int_{0}^{t}\sigma^{2}_{u}\D u-K_{\mathbb{V}}^{0,T}+\frac{1}{\tau}\frac{\tau}{T}\mathbb{E}_{t}\left(\int_{t}^{T}\sigma^{2}_{u}\D u\right)\right)\\
 & =    \E^{-r\tau}\left(\frac{t}{T}\left(\frac{1}{t}\int_{0}^{t}\sigma_{u}^{2}\D u-K_{\mathbb{V}}^{0,T}\right)
+\frac{\tau}{T}\mathbb{E}_{t}\left(\frac{1}{\tau}\int_{t}^{T}\sigma_{u}^{2}\D u-K_{\mathbb{V}}^{0,T}\right)\right).
\end{align*}
But $K_{\mathbb{V}}^{t,T}=\mathbb{E}_{t}\left(\tau^{-1}\int_{t}^{T}\sigma^{2}_{u}du\right)$, and so
$$\Pi_{t}^{T}=\E^{-r\tau}\left(\frac{1}{T}\int_{0}^{t}\sigma_{u}^{2}du-K_{\mathbb{V}}^{0,T}+\frac{\tau}{T}K_{\mathbb{V}}^{t,T}\right).$$
Hence, we just need to calculate the new strike of the variance swap with the remaining maturity $\tau=T-t$. 
A gamma swap looks like a variance swap, but weighs the daily square returns by the price level. Formally speaking, its payoff is worth
$$\mathbb{V}=N\left(T^{-1}\int_{0}^{T}\sigma^{2}_{t}\frac{S_{t}}{S_{0}}\D t-K_{\Gamma}^{0,T}\right),$$
where similarly $K_{\Gamma}^{0,T}$ represents the fair strike of the gamma swap as defined in the contract.
The replication for this option is based on both the It\^o formula and the Carr-Madan formula \cite{CM}.
Let us consider the function $f:\mathbb{R}_+\to\mathbb{R}$ defined by
$$f\left(F_{t}\right):=\E^{rt}\left(F_{t}\log\left(F_{t}/F_{0}\right)-F_{t}+F_{0}\right),$$
where $F:=\left(F_{t}\right){t\geq 0}$ represents the forward price process of $S$. We have $\left(\D F_{t}\right)^2=\sigma_{t}^{2}F_{t}^{2}\D t$, and, by It\^o's formula, we have
\begin{align*}
f\left(F_{T}\right) & = \int_{0}^{T}\left(\partial_t f\right)\D t+\int_{0}^{T}\left(\partial_F f\right)\D F_{t}+\int_{0}^{T}\left(\partial_{FF}^{2} f\right)\sigma_{t}^{2}F_{t}^{2}\D t\\
 & = r\int_{0}^{T}\E^{rt}\left(F_{t}\log\left(F_t/F_0\right)-F_{t}+F_{0}\right)\D t
  + \int_{0}^{T}\E^{rt}\log\left(F_t/F_0\right)\D F_t+\frac{1}{2}\int_{0}^{T}\E^{rt}\sigma_t^2 F_t\D t.
\end{align*}
Hence, the floating leg of the gamma swap can be written as
$$\frac{1}{T}\int_{0}^{T}\sigma_t^2\frac{S_t}{S_0}\D t = \frac{2}{TS_0}\left(f\left(F_T\right)-r\int_{0}^{T}\E^{rt}\left(F_{t}\log\left(F_t/F_0\right)-F_t+F_0\right)\D t 
- \int_{0}^{T}\E^{rt}\log\left(F_t/F_0\right)\D F_t\right),$$
where we used the fact that $S_t=F_{t}\E^{rt}$.
Now, Carr and Madan~\cite{CM} proved that for any continuous function $\phi$ of the forward $F$, we can write
$$
\phi\left(F_{T}\right)  = \phi\left(\kappa\right)+\phi'\left(\kappa\right)\left(\left(F_{T}-\kappa\right)_{+}-\left(\kappa-F_{T}\right)_{+}\right)
 + \int_{0}^{\kappa}\phi''\left(K\right)\left(F_{T}-K\right)_{+}\D K+\int_{\kappa}^{+\infty}\phi''\left(K\right)\left(K-F_{T}\right)_{+}\D K,
$$
where $\kappa>0$ represents a threshold (for example a liquidity threshold in the case of a variance or a gamma swap). 
We now consider the function $\phi:\mathbb{R}_+\to\mathbb{R}$ defined by $\phi(x):=\E^{-rT}f(x)$, where $f$ is defined above. 
Taking $\kappa=F_{0}$, the at-the-money forward spot price, we obtain
$$\phi\left(F_{T}\right)=\int_{0}^{F_{0}}\frac{\D K}{K}\left(K-F_{T}\right)_{+}+\int_{F_{0}}^{+\infty}\frac{\D K}{K}\left(F_{T}-K\right)_{+}.$$
Plugging this equation into the above floating leg of the gamma swap, we eventually have
\begin{align*}
\frac{1}{T}\int_{0}^{T}\sigma^{2}_{t}\frac{S_{t}}{S_{0}}\D t & =\frac{2}{TS_{0}}\E^{rT}\left(\int_{0}^{F_{0}}\frac{1}{K}\left(K-F_{T}\right)_{+}\D K
+\int_{F_{0}}^{+\infty}\frac{1}{K}\left(F_{T}-K\right)_{+}\D K\right)\\
 & -\frac{2}{TS_{0}}\left(r\int_{0}^{T}\E^{rt}\left(F_{t}\log\left(F_{t}/F_{0}\right)-F_{t}+F_{0}\right)\D t
+\int_{0}^{T}\E^{rt}\log\left(\frac{F_{t}}{F_{0}}\right)\D F_{t}\right).
\end{align*}
Hence, a long position in a gamma swap consists of a long continuum of calls and puts weighted by the inverse of the strike, 
rolling a futures position $-2\log\left(F_{t}/F_{0}\right)$ and 
holding a zero-coupon bond, worth $-2r\left(F_{t}\log\left(F_{t}/F_{0}\right)-F_{t}+F_{0}\right)$ at time $t$.

At time $t=0$, the fair value of the gamma swap is hence:
$$
K_{\Gamma}^{0,T} = \mathbb{E}_{0}\left(\frac{1}{T}\int_{0}^{T}\sigma_{t}^{2}\frac{S_{t}}{S_{0}}\D t\right)
 = \frac{2\E^{2rT}}{TS_{0}}\left(\int_{0}^{F_{0}}\frac{\D K}{K}P\left(S_{0},K,T\right)
+\int_{F_{0}}^{+\infty}\frac{\D K}{K}C\left(S_{0},K,T\right)\right).
$$
We can also calculate the price of the gamma swap at time $t=T-\tau$,
$$\frac{1}{T}\int_{0}^{t}\sigma_{u}^{2}\frac{S_{u}}{S_{0}}\D u -K_{\Gamma}^{0,T}
=\frac{1}{T}\int_{0}^{t}\sigma_{u}^{2}\frac{S_{u}}{S_{0}}\D u +\frac{1}{T}\int_{t}^{T}\sigma_{u}^{2}\frac{S_{u}}{S_{0}}\D u-K_{\Gamma}^{0,T}.$$
The first term of the right side of the equation is past, and the two other terms are strikes, hence
$$\mathbb{E}_{t}\left(\frac{1}{T}\int_{0}^{T}\sigma_{u}^{2}\frac{S_{u}}{S_{0}}\D u-K_{\Gamma}^{0,T}\right)
=\frac{1}{T}\int_{0}^{t}\sigma_{u}^{2}\frac{S_{u}}{S_{0}}\D u+\frac{\tau}{T}K_{\Gamma}^{t,T}-K_{\Gamma}^{0,T}.$$

Both variance and gamma swaps provide exposure to volatility. 
However, one of the main difference, from a management point of view, is that variance swaps offer a constant cash gamma, 
whereas gamma swaps provide a constant share gamma, and hence does not require a dynamic reallocation. 
Furthermore, as gamma swaps is weighted by the performance of the underlying stock, it takes into account jumps in it, 
hence traders do not need to cap it, as it is the case for variance swaps (through the use of conditional variance swaps, pp variance swaps, corridor variance swaps).
\section{Correlation trading}
\subsection{Implied correlation}
Consider an index (i.e. a basket) with $n$ stocks. $\sigma_{i}$ represents the volatility of the $i$th stock, $w_{i}$ its weight within the index, 
and $\rho_{i}$ the correlation between stocks $i$ and $j$. If we replicate the index, we create a basket with the following volatility
$$\sigma_{I}^{2}:=\sum_{i=1}^{n}w_{i}^{2}\sigma_{i}^{2}+\sum_{i=1,j\ne i}^{n}w_{i}w_{j}\sigma_{i}\sigma_{j}\rho_{ij}.$$
We can then define the implied correlation in this portfolio, namely an average level of correlation, as
$$\rho_{imp}:=\left(\sum_{i=1,j \ne i}^{n}w_{i}w_{j}\sigma_{i}\sigma_{j}\right)^{-1}\left(\sigma_{I}^{2}-\sum_{i=1}^{n}w_{i}^{2}\sigma_{i}^{2}\right),$$
where $\sigma_{I}$ represents the volatility of the index.
We can rewrite this formula as
$$\rho_{imp}=\left(\sum_{i=1}^{n}\sum_{j>i}\sigma_{i}\sigma_{j}\right)^{-1}\sum_{i=1}^{n}\sum_{j>i}\rho_{ij}w_{i}w_{j}\sigma_{i}\sigma_{j}.$$
In \cite{Bossu1}, Bossu assumed that, under some reasonable conditions, the term $\sum\limits_{i=1}^{n}w_{i}\sigma_{i}^{2}$ is close to zero 
and hence, a good proxy for the implied correlation is
$$\rho_{imp}=\left(\sum_{i=1}^{n}w_{i}\sigma_{i}\right)^{-2}\sigma_{I}^{2}.$$

\subsection{Correlation swaps}
A correlation swap is an instrument similar to a variance swap, 
and pays at maturity the notional multiplied by the difference between the realised correlation and a strike. 
Mathematically speaking, the payoff of such an option is
$$\varPi=-K+\left(\sum_{1\leq i < j\leq n}w_{i}w_{j}\right)^{-1}\sum_{1\leq i < j\leq n}w_{i}w_{j}\rho_{ij}.$$
Similar to the implied correlation above, the realised correlation $\rho$ above can be approximated as
$$\rho=\left(\sum_{i=1}^{n}w_{i}\sigma_{i}\right)^{-2}\sigma_{I}^{2}\approx \left(\sum_{i=1}^{n}w_{i}\sigma_{i}^{2}\right)^{-1}\sigma_{I}^{2},$$
where $\left(\sigma_{I},\sigma_{1},\ldots,\sigma_{n}\right)$ account for realised volatilities. 
We refer to Bossu~\cite{Bossu1} and the corresponding presentation for the details of this approximation 
(which is, in fact, a lower bound, thanks to Jensen's inequality). 
Hence, the realised correlation can be seen as the ratio between two traded products, through variance swaps, or variance dispersion trades. 
Based on this proxy, Bossu~\cite{Bossu1} proved that the correlation swap can be dynamically quasi-replicated by a variance dispersion trade, 
and that the P\&L of a variance dispersion trading is worth $\sum\limits_{i=1}^{n}w_{i}\sigma_{i}^{2}\left(1-\rho\right)$, 
where $\rho$ represents the realised correlation.

Though this result is really nice, several issues need to be pointed out: 
first, liquidity is not enough on all markets for variance swaps, neither for every index and its components; 
then this model does not specify the form of the volatility. 
Indeed, it does not take into account the possible random moves in the volatility, namely a volatility of volatility parameter.
\section{Dispersion trading}
\subsection{P\&L of a delta-hedged portfolio, with constant volatility}
We here consider an option $V_{t}$, valued at time $t\geq 0$ written on the asset $S$. 
The hedged portfolio consists of being short the option and long $\delta_t$ of the stock price, resulting in a certain amount of cash. 
Namely, the P\&L variation of the portfolio $\varPi$ at time $t$ is worth
$$\Delta \varPi_{t}=\Delta V_{t}-\delta \Delta S_{t}+\left(\delta S_{t}-V_{t}\right)r\Delta t.$$
The first part corresponds to the price variation of the option, the second one to the stock price movements, 
of which we hold $\delta$ units, and the third part is the risk-free return of the amount of cash to make the portfolio have zero value. 
Now, Taylor expanding the option price gives
$$\Delta V_{t}=\delta_t \Delta S+\frac{1}{2}\Gamma_t \left(\Delta S\right)^{2}+\theta_t \Delta t,$$
where $\theta_t:=\partial_t V_t$ and $\Gamma_t:=\partial_{SS}^2 V_t$, hence, the P\&L variation reads
$$\Delta \varPi_{t}=\delta_t\Delta S+\frac{1}{2}\Gamma_t\left(\Delta S\right)^{2}+\theta_t\Delta t-\delta_t \Delta S_{t}+\left(\delta_t S_{t}-V_{t}\right)r\Delta t.$$
Moreover, as the option price follows the Black \& Scholes PDE 
$\theta_t+rS_{t}\delta_t+\frac{1}{2}\sigma^{2}S_{t}^{2}\Gamma_t=rV_{t}$, 
we obtain the final P\&L for the portfolio on $\left(t,t+\D t\right)$ : 
$$
P\&L_{\left(t,t+\D t\right)}=\frac{1}{2}\Gamma_t S_{t}^{2}\left(\left(\frac{\D S_{t}}{S_{t}}\right)^{2}-\sigma_{t}^{2}\D t\right).
$$
\subsection{P\&L of a delta-hedged portfolio, with time-running volatility}
We here consider the P\&L of a trader who holds an option and delta-hedges it with the underlying stock. 
As we do wish to analyse the volatility risk, we stay in this incomplete market, 
as opposed to traditional stochastic volatility option pricing framework. 
The dynamics for the stock price is now $\D S_{t}/S_t=\mu \D t+\sigma_{t}\D W_{t}$, with $\mu\in\mathbb{R}$. 
For the volatility process, we assume the following dynamics: 
$\D \sigma_{t}=\mu_{\sigma,t}\D t+\xi\sigma_{t}\D W_{t}^{\sigma}$, with $\sigma_0, \xi>0$, $\mu_{\sigma}\in\mathbb{R}$ and $\D <W,W^{\sigma}>_{t}=\rho \D t$. 
As before, the P\&L of the trader on the period $\left(t,t+\D t\right)$ is 
$$\Delta \varPi_{t}=\Delta V_{t}-\delta_t \Delta S_{t}+\left(\delta_t S_{t}-V_{t}\right)r\Delta t.$$
We now use a Taylor expansion of $\Delta V$ with respect to the time, the stock and the volatility
$$\Delta V_{t}=\theta_t \D t+\partial_S V_t \Delta S_{t}+\partial_{\sigma} V_t\Delta \sigma_t+\frac{1}{2}\left(\partial_{SS}^{2}V_t\left(\Delta S_t\right)^{2}
+\partial_{\sigma\sigma}^{2}V_t\left(\Delta \sigma_t\right)^{2}+2\partial_{S\sigma}^{2}V_t\Delta S_t\Delta \sigma_t\right).$$
Now, in the P\&L formula, we can replace the $rV_{t}dt$ term by its value given in the Black-Scholes PDE, calculated with the implied volatility. 
Indeed, this is the very volatility that had to be input to determine the amount of cash to lock the position. Hence, we obtain
\begin{align}\label{eq:PLrunningVol}
P\&L_t= & \theta_t \D t+\partial_S V_t\D S_{t}+\partial_{\sigma} V_t\D\sigma_t+\frac{1}{2}\left(\partial_{SS}^{2}V_t\left(\D S_t\right)^{2}
+\partial_{\sigma\sigma}^{2}V_t\left(\D\sigma\right)^{2}+2\partial_{S\sigma}^{2}V_t\D S_t \D \sigma_t\right)\nonumber\\
 & -\delta_t \D S_{t}+r\delta_t S_{t}\D t-\left(\theta_t+\frac{1}{2}\tilde{\sigma}_t^{2}S_{t}^{2}\partial_{SS}^{2}V_t+rS_{t}\partial_S V_t\right)\D t,
\end{align}
where $\tilde{\sigma}_t$ represents the implied volatility of the option. We can rewrite \eqref{eq:PLrunningVol} as
$$P\&L_t=\frac{1}{2}\Gamma S_{t}^{2}\left(\left(\frac{\D S_{t}}{S_{t}}\right)^{2}-\tilde{\sigma}_t^{2}\D t\right)+\partial_{\sigma} V_t\D\sigma
+\frac{1}{2}\partial_{\sigma\sigma}^{2} V_t\left(\D\sigma_t\right)^{2}+\partial_{SS}^{2} V_tS_{t}\sigma_{t}\D W_{t}\D\sigma_t$$
In trading terms, this can be expressed as
\begin{equation}\label{eq:eq2}
P\&L_t=\frac{1}{2}\Gamma S_{t}^{2}\left(\left(\frac{\D S_{t}}{S_{t}}\right)^{2}-\tilde{\sigma}_t^{2}\D t\right)+\text{Vega }\D\sigma
+\frac{1}{2}\text{Volga }\xi^{2}\sigma_{t}^{2}\D t+\text{Vanna }\sigma_{t}S_{t}\rho\xi\D t,
\end{equation}
where
$\text{Vega}=\partial_{\sigma} V$, 
$\text{Volga}=\partial_{\sigma\sigma}^{2} V$, 
$\text{Vanna}=\partial_{S\sigma}^{2} V$, 
$\rho$ represents the correlation between the stock price and the volatility and 
$\xi$ is volatility of volatility as defined above.

\subsection{Delta-hedged dispersion trades with $\D\sigma=0$}
We consider the dispersion trade as being short the index option and long the stock options. We also consider it delta-hedged. 
The P\&L of a delta-hedged option $\varPi$ in the Black-Scholes framework is
$$P\&L=\theta_t\left(\left(\frac{\D S}{S\sigma_t\sqrt{\D t}}\right)^{2}-1\right).$$
The term $n:=\D S_t / \left(S_t\sigma_t\sqrt{\D t}\right)$ represents the standardised move of the underlying on the considered period. 
Let us now consider an index $I$ composed by $n$ stocks $\left(S_{i}\right)_{i=1,\ldots,n}$. For convenience, we drop the time index on the stock price processes.
We first develop the P\&L of a long position in the index, in terms of its constituents, 
then decompose it into idiosyncratic risk and systematic risk. Let us denote
$n_{i}:=\D S_{i} / \left(S_{i}\sigma_{i}\sqrt{\D t}\right)$ the standardised move of the $i$th stock, 
$n_{I}:=\D S_{I} / \left(S_{I}\sigma_{I}\sqrt{\D t}\right)$ the standardised move of the index, 
$p_{i}$ the number of shares $i$ in the index,
$w_{i}$ the weight of share $i$ in the index,
$\sigma_{i}$ the volatility of stock $i$,
$\sigma_{I}$ the volatility of the index,
$\rho_{ij}$ the correlation between stocks $i$ and $j$,
$\theta_{i}$ the theta of the option written on stock $i$,
$\theta_{I}$ the theta of the option written on the index.
The P\&L of the index can hence be written as : 
\begin{align*}
P\&L_t & =\theta_{I}\left(n_{I}^{2}-1\right)=
 \theta_{I}\left(\left(\sum_{i=1}^{n}w_{i}n_{i}\frac{\sigma_{i}}{\sigma_{I}}\right)^{2}-1\right)\\
 & = \theta_{I}\left(\sum_{i=1}^{n}\left(w_{i}n_{i}\frac{\sigma_{i}}{\sigma_{I}}\right)^{2}+\sum_{i \ne j}\frac{w_{i}w_{j}\sigma_{i}\sigma_{j}}{\sigma_{I}}n_{i}n_{j}-1\right)\\
 & = \theta_{I}\sum_{i}^{n}\frac{w_{i}^{2}\sigma_{i}^{2}}{\sigma_{I}^{2}}\left(n_{i}^{2}-1\right)+\theta_{I}\sum_{i\ne j}\frac{\sigma_{i}\sigma_{j}}{\sigma_{I}^{2}}w_{i}w_{j}\left(n_{i}n_{j}-\rho_{ij}\right), 
\end{align*}
where we used the fact that
$$
n_{I} := \frac{\D I}{I\sigma_{I}\sqrt{\D t}}
= \left(\sigma_{I}\sqrt{\D t}\right)^{-1}\frac{\sum_{i=1}^{n}p_{i}\D S_{i}}{\sum_{j=1}^{n}p_{j}S_{j}}
= \sigma_{I}^{-1}\sum_{i=1}^{n}\frac{p_{i}S_{i}}{\sum_{j=1}^{n}p_{j}S_{j}}\sigma_{i}\frac{\D S_{i}}{\sigma_{i} S_{i}\sqrt{\D t}}
= \sum_{i=1}^{n}w_{i}\frac{\sigma_{i}}{\sigma_{I}}n_{i},
$$
and
$$\sigma_{I}^{2}=\sum_{i=1}^{n}\left(w_{i}\sigma_{i}\right)^{2}n_{i}^{2}+\sum_{i\ne j}w_{i}w_{j}\sigma_{i}\sigma_{j}\rho_{ij}.$$
Hence, the dispersion trade, namely shorting the index option and being long the options on the stocks has the following P\&L 
$$
P\&L = \sum_{i=1}^{n}P\&L_{i}-P\&L_{I}
 = \sum_{i=1}^{n}\theta_{i}\left(n_{i}^{2}-1\right)+\theta_{I}\left(n_{I}^{2}-1\right).
$$
The short and long position in the options will be reflected in the sign of the $\left(\theta_{1},\ldots,\theta_{n},\theta_{I}\right)$. 
More precisely, a long position will mean a positive $\theta$ whereas a short position will have a negative $\theta$.
\subsection{Weighting schemes for dispersion trading}
Here above, when considering the weights of the stocks in the index, we did not specify what they precisely were. 
In fact, when building a dispersion trade, one faces two problems : first, which stocks to pick? 
Then, how to weight them? As there may lack liquidity on some stocks, the trader will not take into account all the components of the index. 
Thus, he would rather select those that show great characteristics and liquidity. From his point of view, he can build several weighting strategies:
\begin{itemize}
\item Vega-hedging weighting: 
the trader will build his dispersion such that the vega of the index equals the sum of the vegas of the constituents. 
Hence, this will immune him against short moves in the volatility.

\item Gamma-hedging weighting: 
The gamma of the index is worth the sum of the gammas of the components. 
As the portfolio is already delta-hedged, this weighting scheme protects the trader against any move in the stocks, but leaves him with a vega position.

\item Theta-hedging weighting: 
this strategy is rather different from the previous two, as it will result in both a short vega as well as a short gamma position.
\end{itemize}

\section{Correlation swaps vs dispersion trades}
We here focus of the core topic of our paper, namely the difference between the strike of a correlation swap and the implied correlation obtained through dispersion trading. 
Empirical proofs do observe a spread of approximately ten basis points between the strike of a correlation swap and the dispersion implied correlation. 
We first write down the relation between a dispersion trade through variance swaps and a correlation swap; 
thanks to this relation, we analyse the influence of the dynamics of the volatility on this very spread. 
In the whole section, we will consider that the nominal of the index variance swap is equal to $1$.

\subsection{Analytical formula for the spread}
We keep the previous notations, namely an index $I$, with implied volatility $\sigma_{I}$, 
realised volatility $\hat{\sigma}_{I}$ composed of $n$ stocks with characteristics 
$\left(\sigma_{i},\hat{\sigma_{i}},w_{i}\right)_{i=1,\ldots n}$. 
The implied and the realised correlation are obtained as previously,
$$\sigma_{I}^{2}=\sum_{i=1}^{n}w_{i}\sigma_{i}^{2}+\sum_{i\ne j}w_{i}w_{j}\rho\sigma_{i}\sigma_{j},$$
and
$$\hat{\sigma}_{I}^{2}=\sum_{i=1}^{n}w_{i}\hat{\sigma}_{i}^{2}+\sum_{i\ne j}w_{i}w_{j}\hat{\rho}\hat{\sigma}_{i}\hat{\sigma}_{j}.$$
If we subtract these two equalities, we obtain
\begin{align*}
\hat{\sigma}_{I}^{2}-\sigma_{I}^{2} & = \sum_{i=1}^{n}w_{i}^{2}\left(\hat{\sigma}_{I}^{2}-\sigma_{i}^{2}\right)+\sum_{i\ne j}w_{i}w_{j}\left(\hat{\sigma_{i}}\hat{\sigma_{j}}\hat{\rho}-\sigma_{i}\sigma_{j}\rho\right)\\
 & = \sum_{i=1}^{n}w_{i}^{2}\left(\hat{\sigma}_{I}^{2}-\sigma_{i}^{2}\right)+\sum_{i\ne j}w_{i}w_{j}\Big(\sigma_{i}\sigma_{j}\left(\hat{\rho}-\rho\right) +\left(\hat{\sigma_{i}}\hat{\sigma_{j}}-\sigma_{i}\sigma_{j}\right)\hat{\rho}\Big)\\
 & = \sum_{i=1}^{n}w_{i}^{2}\left(\hat{\sigma}_{I}^{2}-\sigma_{i}^{2}\right)+\sum_{i\ne j}w_{i}w_{j}\sigma_{i}\sigma_{j}\left(\hat{\rho}-\rho\right)+\sum_{i\ne j}\hat{\rho}\left(\hat{\sigma_{i}}\hat{\sigma_{j}}-\sigma_{i}\sigma_{j}\right),
\end{align*}
and hence
$$\left(\sum_{i=1}^{n}w_{i}^{2}\left(\hat{\sigma}_{I}^{2}-\sigma_{i}^{2}\right)-\left(\hat{\sigma}_{I}^{2}-\sigma_{I}^{2}\right)\right)-\sum_{i\ne j}w_{i}w_{j}\sigma_{i}\sigma_{j}\left(\hat{\rho}-\rho\right)=\sum_{i\ne j}\hat{\rho}\left(\hat{\sigma_{i}}\hat{\sigma_{j}}-\sigma_{i}\sigma_{j}\right).$$
From a financial point of view, the above formula evaluates the P\&L of a position consisting of being short a dispersion trade 
through variance swaps (short the index variance swap and long the components' variance swaps) and long a correlation swap. 
The right-hand side, the P\&L, is the spread we are considering.
\subsection{Gamma P\&L of the dispersion trade}
We here only consider the gamma part of the P\&L of the variance swap of the index. We have (as in section 4.3)
\begin{align*}
P\&L_{I}^{\Gamma} & = \frac{1}{2}\Gamma_{I}I_{t}^{2}\left(\left(\D I_{t} / I_{t}\right)^{2}-\sigma_{I}^{2}\D t\right)
 = \frac{1}{2}\Gamma_{I}I_{t}^{2}\left(\left(\sum_{i=1}^{n}w_{i}\frac{\D S_{i}}{S_{i}}\right)^{2}-\left(\sum_{i=1}^{n}w_{i}^{2}\sigma_{i}^{2}+\sum_{i\ne j}w_{i}w_{j}\sigma_{i}\sigma_{j}\rho_{ij}\right)\D t\right)\\
 & = \frac{1}{2}\Gamma_{I}I_{t}^{2}\left(\sum_{i=1}^{n}w_{i}^{2}\left(\frac{\D S_{i}}{S_{i}}\right)^{2}+\sum_{i\ne j}w_{i}w_{j}\frac{\D S_{i}\D S_{j}}{S_{i}S_{j}}-\sum_{i=1}^{n}w_{i}^{2}\sigma_{i}^{2}-\sum_{i\ne j}w_{i}w_{j}\sigma_{i}\sigma_{j}\rho_{ij}\right)\\
 & = \frac{1}{2}\Gamma_{I}I_{t}^{2}\left(\sum_{i=1}^{n}w_{i}^{2}\left(\left(\frac{\D S_{i}}{S_{i}}\right)^{2}-\sigma_{i}^{2}\D t\right)+\sum_{i\ne j}w_{i}w_{j}\sigma_{i}\sigma_{j}\left(\frac{\D S_{i}\D S_{j}}{S_{i}S_{j}\sigma_{i}\sigma_{j}}-\rho_{ij}\D t\right)\right).
\end{align*}
Let us make a pause to analyse this formula. As before, in the context of dispersion, 
we assume that the correlations $\rho_{ij}$ between the components are all equal to an average one $\rho$. 
Furthermore, as this correlation is the one that makes the implied variance of the index and the implied variance of the weighted sum of the components equal, 
it exactly represents the implied correlation. 
Then $\D S_{i}\D S_{j} / \left(S_{i}S_{j}\right)$ accounts for the instantaneous realised covariance 
between the two stocks $S_{i}$ and $S_{j}$, and hence $\D S_{i}\D S_{j} / \left(S_{i}S_{j}\sigma_{i}\sigma_{j}\right)$ is precisely 
the instantaneous realised correlation between the two stocks. Again, we assume it is the same for all pairs of stocks, 
and we note it $\hat{\rho}$. Then, we can replace the weights $w_{i}=p_{i}S_{i} / I$. We therefore obtain
$$P\&L_{I}^{\Gamma}=\frac{1}{2}\Gamma_{I}\left(\sum_{i=1}^{n}p_{i}^{2}S_{i}^{2}\left(\left(\frac{\D S_{i}}{S_{i}}\right)-\sigma_{i}^{2}\D t\right)+I_{t}^{2}\sum_{i\ne j}w_{i}w_{j}\sigma_{i}\sigma_{j}\left(\hat{\rho}-\rho\right)\D t\right).$$
Suppose we consider a position in a dispersion trade with variance swaps, 
where $\alpha_{i}$ represents the proportion of variance swaps for the $i$th stock, the gamma P\&L is then worth
$$
\sum_{i=1}^{n}\alpha_{i}P\&L_{i}^{\Gamma}-P\&L_{I}^{\Gamma} 
= \sum_{i=1}^{n}\frac{1}{2}S_{i}^{2}\left(\left(\frac{\D S_{i}}{S_{i}}\right)-\sigma_{i}^{2}\D t\right)\left(\alpha_{i}\Gamma_{i}-p_{i}^{2}\Gamma_{I}\right)
+ \frac{1}{2}\Gamma_{I}\sum_{i\ne j}p_{i}p_{j}\sigma_{i}\sigma_{j}S_{i}S_{j}\left(\rho-\hat{\rho}\right)\D t.
$$
The P\&L of the dispersion trade is hence equal to the sum of a spread between the implied and the realised correlation 
over a period of time $\left(t,t+\D t\right)$ (pure correlation exposure) and a volatility exposure.
Now, we recall that the gamma of a variance swap for a maturity $T$ is $\Gamma=2/\left(T S^{2}\right)$. 
Hence, we can rewrite the gamma P\&L for the index variance swap as
$$P\&L_{I}^{\Gamma}=\sum_{i=1}^{n}\frac{1}{T}p_{i}^{2}\frac{S_{i}^{2}}{I_{t}^{2}}\left(\left(\frac{\D S_{i}}{S_{i}}\right)
-\sigma_{i}^{2}\D t\right)+\frac{1}{T I_{t}^{2}}\sum_{i\ne j}p_{i}p_{j}\sigma_{i}\sigma_{j}S_{i}S_{j}\left(\hat{\rho}-\rho\right)\D t,$$
which implies
$$
\sum_{i=1}^{n}\alpha_{i}P\&L_{i}^{\Gamma}-P\&L_{I}^{\Gamma} 
= \frac{1}{T}\sum_{i=1}^{n}\left(\left(\frac{\D S_{i}}{S_{i}}\right)-\sigma_{i}^{2}\D t\right)\left(\alpha_{i}-p_{i}^{2}\frac{S_{i}^{2}}{I_{t}^{2}}\right)\\
  +\frac{1}{T}\sum_{i\ne j}w_{i}w_{j}\sigma_{i}\sigma_{j}\left(\rho-\hat{\rho}\right)\D t,
$$
and hence
$$
\sum_{i=1}^{n}\alpha_{i}P\&L_{i}^{\Gamma}-P\&L_{I}^{\Gamma} = \frac{1}{T}\sum_{i=1}^{n}\left(\left(\frac{\D S_{i}}{S_{i}}\right)^{2}-\sigma_{i}^{2}\D t\right)\left(\alpha_{i}-w_{i}^{2}\right)\\
  +\frac{1}{T}\sum_{i\ne j}w_{i}w_{j}\sigma_{i}\sigma_{j}\left(\rho-\hat{\rho}\right)\D t.
$$
The sum that multiplies the correlation spread does not depend on the correlation, but only on the components of the index. 
Hence, we can define $\beta^{V}:=T^{-1}\sum\limits_{i\ne j}w_{i}w_{j}\sigma_{i}\sigma_{j}$ and eventually write
\begin{equation}
P\&L_{\text{Disp}}^{\Gamma}=\frac{1}{T}\sum_{i=1}^{n}\left(\left(\frac{\D S_{i}}{S_{i}}\right)^{2}-\sigma_{i}^{2}\D t\right)\left(\alpha_{i}-w_{i}^{2}\right)+\beta^{V}\left(\rho-\hat{\rho}\right)\D t.
\end{equation}
If we take $\alpha_{i}=w_{i}^{2}$, then we see that the gamma P\&L of the dispersion trade is exactly the spread 
between implied and realised correlation, multiplied by a factor $\beta$ which corresponds to a weighted average variance of the components of the index
\begin{equation}\label{eq:eq4}
P\&L_{\text{Disp}}^{\Gamma}=\beta^{V}\left(\rho-\hat{\rho}\right)\D t.
\end{equation}

In fact, as we will analyse it later, this weighting scheme is not used. However, this approximation (considering that the gamma P\&L is pure correlation exposure) is quite fair, and we will measure the induced error further in this paper.
\subsection{Total P\&L of the dispersion trade}
In the previous subsection, we proved that the gamma P\&L of a dispersion trade is exactly a correlation P\&L
. Hence, the observed difference between the implied correlation of a dispersion trade and the strike of the correlation swap with the same characteristics 
(about ten basis points) is precisely due to the volatility terms, namely the combined effects of the vega, the volga (vomma) and vanna.
Using (\ref{eq:eq2}) and (\ref{eq:eq4}), we can now write
$$P\&L_{\text{Disp}}=P\&L_{\text{Disp}}^{\Gamma}+P\&L_{\text{Disp}}^{\text{Vol}},$$
Where the $P\&L^{\Gamma}$ contains the correlation exposure, and the $P\&L^{\text{Vol}}$ contains all the vegas, volgas and vannas. More precisely,
\begin{align*}
P\&L_{\text{Disp}}^{\text{Vol}} & = \sum_{i=1}^{n}\alpha_{i}\left(\text{Vega}_{i}\D \sigma_{i}+\frac{1}{2}\text{Volga}_{i}\xi_{i}^{2}\sigma_{i}^{2}\D t+\text{Vanna}_{i}\sigma_{i}S_{i}\rho_{i}\xi_{i} \D t\right)\\
 & -\left(\text{Vega}_{I}\D \sigma_{I}+\frac{1}{2}\text{Volga}_{i}\xi_{I}^{2}\sigma_{I}^{2}\D t+\text{Vanna}_{I}\sigma_{I}I\rho_{I}\xi_{I}\D t\right).
\end{align*}
When replacing the greeks by their values for a variance swap, we obtain (the vanna being null):
$$
P\&L_{\text{Disp}} 
= P\&L_{\text{Disp}}^{\Gamma}+2\frac{\tau}{T}\left(\left(\sum_{i=1}^{n}\alpha_{i}\sigma_{i}^{\text{imp}}\D \sigma_{i}\right)-\sigma^{\text{imp}}\D \sigma\right)\D t
  +\frac{\tau}{T}\left(\left(\sum_{i=1}^{n}\alpha_{i}\xi_{i}^{2}\sigma_{i,t}^{2}\right)-\xi_{I}^{2}\sigma_{I}^{2}\right)\D t.
$$
\subsection{P\&L with different weighting schemes}
\noindent
In the following weighting schemes strategy, we will consider that the gamma P\&L of the dispersion trade is pure correlation exposure, 
hence respects \eqref{eq:eq4}. 
Concerning the notations, $\alpha_{i}$ is still the proportion of variance swaps of the $i$th stock ($\alpha_{i}=N_{i}/N_{I}$), 
and we consider $N_{I}=1$, $w_{i}$ the weight of stock $i$ in the index and $N_{i}$ represents the notional of the $i$th variance swap. 
We also do not write the negative signs for the greeks ; therefore, when writing the greek of a product, 
one has to bear in mind that its sign depend on the position the trader has on this very product.

\textbf{Vega flat Strategy}\\
In this strategy, the vega notional of the index variance swap is equal to the sum of the vega notionals of the components : 
$N_{i}\Upsilon_{\sigma,i}=N_{I}\Upsilon_{\sigma,I}w_{i}$, and 
$\alpha_{i}=\sigma_{I}w_{i}/\sigma_i$.
The vegas of the P\&L disappear and we are left with
$$P\&L_{\text{Disp}}=P\&L_{\text{Disp}}^{\Gamma}+\frac{\tau}{T}\left(\left(\sum_{i=1}^{n}\alpha_{i}\xi_{i}^{2}\sigma_{i,t}^{2}\right)-\xi_{I}^{2}\sigma_{I}^{2}\right)\D t,$$
with the above mentioned approximation, we therefore have
$$P\&L_{\text{Disp}}=\beta^{V}\left(\rho-\hat{\rho}\right)dt+\frac{\tau}{T}\left(\left(\sum_{i=1}^{n}\frac{\sigma_{I}}{\sigma_{i}}w_{i}\xi_{i}^{2}\sigma_{i,t}^{2}\right)-\xi_{I}^{2}\sigma_{I}^{2}\right)\D t.$$
Now, the error due to the approximation in the gamma P\&L is worth 
$$\sum_{i=1}^{n}\left(\left(\frac{\D S_{i}}{S_{i}}\right)^{2}-\sigma_{i}^{2}\D t\right)\left(\alpha_{i}-w_{i}^{2}\right).$$ 
We now focus on the $\left(\alpha_{i}-w_{i}^{2}\right)$ part. Here we have
$$\alpha_{i}-w_{i}^{2}=w_{i}\frac{\sigma_{I}}{\sigma_{i}}-w_{i}^{2}=w_{i}^{2}\left(\frac{\sigma_{I}}{w_{i}\sigma_{i}}-1\right),$$
which is indeed very close to $0$. From a very theoretical point of view, this formula tells us that the observed difference 
between the strike of a correlation swap and the implied correlation via variance swaps dispersion trades can be simply explained by the volga of the dispersion trade, 
hence, by the volatility of volatility terms.

\textbf{Vega weighted flat strategy}\\
In this strategy, we have : 
$N_{i}\Upsilon_{\sigma,i}=\left(\sum_{j=1}^{n}w_{j}\sigma_{j}\right)^{-1}N_{I}\Upsilon_{\sigma,I}w_{i}\sigma_{I}$, 
and 
$\alpha_{i}=\left(\sum_{j=1}^{n}w_{j}\sigma_{j}\right)^{-1}w_{i}\sigma_{I}\sigma_{I} / \sigma_{i}$. 
We are left with
$$P\&L_{\text{Disp}}=P\&L_{\text{Disp}}^{\Gamma}+\frac{\tau}{T}\left(\left(\sum_{i=1}^{n}\alpha_{i}\xi_{i}^{2}\sigma_{i,t}^{2}\right)-\xi_{I}^{2}\sigma_{I}^{2}\right)\D t+\frac{\tau}{T}\left(\left(\sum_{i=1}^{n}\alpha_{i}\xi_{i}\sigma_{i,t}^{2}\right)-\xi_{I}^{2}\sigma_{I}^{2}\right)\D t,$$
with the above mentioned approximation, we therefore have : 
\begin{align*}
P\&L_{\text{Disp}} & = \beta^{V}\left(\rho-\hat{\rho}\right)dt
+\frac{\tau}{T}\left(\left(\sum_{i=1}^{n}\frac{\sigma_{I}}{\sigma_{i}}w_{i}\sigma_{I}\left(\sum_{j=1}^{n}w_{j}\sigma_{j}\right)^{-1}\xi_{i}^{2}\sigma_{i,t}^{2}\right)
-\xi_{I}^{2}\sigma_{I}^{2}\right)\D t\\
 & + \frac{\tau}{T}\left(\left(\sum_{i=1}^{n}\frac{\sigma_{I}}{\sigma_{i}}w_{i}\sigma_{I}\left(\sum_{j=1}^{n}w_{j}\sigma_{j}\right)^{-1}\xi_{i}^{2}\sigma_{i,t}^{2}\right)-\xi_{I}^{2}\sigma_{I}^{2}\right)\D t.
\end{align*}
The error due to the approximation in the gamma P\&L is then worth 
$$\alpha_{i}-w_{i}^{2}=w_{i}\frac{\sigma_{I}}{\sigma_{i}}\sigma_{I}\left(\sum_{j=1}^{n}w_{j}\sigma_{j}\right)^{-1}-w_{i}^{2}
=w_{i}^{2}\left(\frac{\sigma_{I}}{w_{i}\sigma_{i}}\sigma_{I}\left(\sum_{j=1}^{n}w_{j}\sigma_{j}\right)^{-1}-1\right),$$
which is indeed very close to $0$. From a very theoretical point of view, this formula tells us that the observed difference 
between the strike of a correlation swap and the implied correlation via variance swaps dispersion trades can be simply explained 
by the volga of the dispersion trade, hence, by the volatility of volatility terms.

\textbf{Theta/Gamma flat Strategy}\\
Suppose we want to get rid of the gamma P\&L of the dispersion. Recalling that it is worth
$$P\&L_{\text{Disp}}^{\Gamma}=
\frac{1}{2}\sum_{i=1}^{n}\alpha_{i}\Gamma_{i}S_{i}^{2}\left(\left(\frac{\D S_{i}}{S_{i}}\right)^{2}-\sigma_{i}^{2}\D t\right)-\frac{1}{2}\Gamma_{I}I_{t}^{2}\left(\left(\frac{\D I_{t}}{I_{t}}\right)^{2}-\sigma_{I}^{2}\D t\right).$$
We thus need to set 
$$\alpha_{i}=\frac{\Gamma_{I}I_{t}^{2}\left(\left(\frac{\D I_{t}}{I_{t}}\right)^{2}-\sigma_{I}^{2}\D t\right)}{\sum\limits_{i=1}^{n}\Gamma_{i}S_{i}^{2}\left(\left(\frac{\D S_{i}}{S_{i}}\right)^{2}-\sigma_{i}^{2}\D t\right)},$$
and replacing the $\Gamma_i$ and $\Gamma_I$ by their values, we get
$$\alpha_{i}=\frac{\left(\D I_{t}/I_t\right)^{2}-\sigma_{I}^{2}\D t}{\sum\limits_{i=1}^{n}\left(\left(\D S_{i}/S_{i}\right)^{2}-\sigma_{i}^{2}\D t\right)}.$$
On a very short period, we almost have $\left(\D S/S\right)^{2}=0$, and hence this is also a theta flat strategy (with no interest rate. 
Actually the difference between the gamma and the theta flat strategies is the difference between the risk-free rate 
and the return of the stocks over the period we consider). 
Furthermore, the dispersion trade is fully exposed to moves in volatility, namely through the vegas and the vannas of the variance swaps.
\section{Conclusion}
We have here dealt with dispersion trading, and we showed the P\&L of such a strategy, 
considering both variance swaps and gamma swaps. 
The first one are particularly appealing because of their greeks, which enable us to have a clear vision of our exposure. 
The main result of our paper is that we proved that the observed spread between implied correlation through variance swaps 
dispersion trades and fair values of correlation swaps is totally explained by a vol of vol parameter. 
We also developed results for gamma swaps dispersion trades and different weighting schemes, one of them, the vega flat weighting strategy, being an arbitrage bound. 
This also gives us a way of estimating the volatility of volatility parameter, based on the observed prices of variance and correlation swaps.

This work could be analysed deeper when considering third-generation exotic product such as corridor variance swaps or up variance swaps. They indeed allow investor to bet on future realised variance at a lower cost. Similar results should be found, but with less elegant formulas, as the stock price - just like for gamma swaps - will have to be taken into account.

\appendix
\section{Greeks of a variance swap}
Let $T>0$ represent the maturity of the variance swap and $t$ the valuation date. We denote $t:=T-\tau$, and the following greeks are straightforward,
\begin{align*}
\varDelta=2 T^{-1}\left(S_{*}^{-1}-S_{t}^{-1}\right), & & \Gamma=2 S_{t}^{-2}/T, & & \Theta=-\sigma^{2}/T,\\
\Upsilon_{\sigma^{2}}=\tau/T, & & \partial_S \Gamma=-4 S_{t}^{-3}/T, & & \partial_S V_{\sigma^{2}}=0, \\
\partial_{\tau} \Gamma=0, & & \partial_{\tau} V_{\sigma^{2}}=T^{-1}, & & \Upsilon_{\sigma}=2\sigma\tau / T.
\end{align*}

\section{Greeks of a gamma swap}
Let $T>0$ represent the maturity of the variance swap and $t$ the valuation date. We denote $t:=T-\tau$, and the following greeks are as follows,
\begin{align*}	
\Upsilon_{\sigma}=2T^{-1}\tau\sigma\E^{r\left(T+\tau\right)}S_{t}/S_0, & & 
\Psi=2T^{-1}\tau\sigma\E^{r\left(T+\tau\right)}/S_0,\\
\Lambda_{\sigma}=2T^{-1}\tau\E^{r\left(T+\tau\right)}S_{t}/S_0, & & 
\Gamma=\frac{2}{S_0 S_{t}T}\E^{r\left(T+\tau\right)}.
\end{align*}	

We prove below the value for the vega $\Upsilon_\sigma$ and the gamma $\Gamma$ of the gamma swap. 
we know that the value of a gamma swap at inception is worth
$$\mathbb{E}_{0}\left(\frac{1}{T}\int_{0}^{T}\sigma_{t}^{2}\frac{S_{t}}{S_{0}}\D t\right)
=\frac{2\E^{2rT}}{TS_{0}}\left(\int_{0}^{F_{0}}\frac{\D K}{K}P\left(S_{0},K\right)
+\int_{F_{0}}^{+\infty}\frac{\D K}{K}C\left(S_{0},K\right)\right).$$
Its vega at inception is then
\begin{align*}
\Upsilon_{\sigma}^{\Gamma} & = \frac{2\E^{2rT}}{TS_{0}}\int_{0}^{+\infty}\frac{\D K}{K}\frac{\partial O_{K}}{\partial \sigma}
 = \frac{2\E^{2rT}}{TS_{0}}\int_{0}^{+\infty}\frac{\D K}{K}S_{0}\sqrt{T}\phi\left(d_{1}\right)\\
 & = \frac{2\E^{2rT}\sqrt{T}}{TS_{0}}S_{t}\int_{0}^{+\infty}\frac{\D K}{K\sqrt{2\pi}}\exp\left(-\frac{1}{2\sigma^{2}T}\left(\log\left(S_{0}/K\right)+\left(r+\sigma^{2}/2\right)T\right)^{2}\right)\\
 & = \frac{2\E^{2rT}\sqrt{T}}{TS_{0}}S_{0}\int_{0}^{+\infty}\frac{\D K}{K\sqrt{2\pi}}
\exp\left(-\frac{1}{2\sigma^{2}T}\left(\log\left(S_0\right)-\left(\log\left(K\right)+\left(r+\sigma^{2}/2\right)T\right)\right)^{2}\right),
\end{align*}
where we used the fact that the vega of a call option is equal to the vega of a put option.
Let us do the following change of variable
$x=\left(\log\left(S_0\right)-\left(\log\left(K\right)+\left(r+\frac{\sigma^{2}}{2}\right)T\right)\right)/\left(\sigma\sqrt{T}\right)$. 
We then have
$$
\Upsilon_{\sigma}^{\Gamma} = \frac{2 T\sigma\E^{2rT}}{S_{0}T}S_{0}\int_{-\infty}^{+\infty}\frac{\D x}{\sqrt{2\pi}}\exp\left(-x^{2}/2\right) = 2\sigma \E^{2rT}.
$$
Moreover, at time $t=T-\tau$, we have
$$\mathbb{E}_{t}\left(\frac{1}{T}\int_{0}^{T}\sigma_{u}^{2}\frac{S_{u}}{S_{0}}du\right)=\frac{1}{T}\int_{0}^{t}\sigma_{u}^{2}\frac{S_{u}}{S_{0}}du+\frac{\tau}{T}\frac{S_{t}}{S_{0}}K_{\Gamma}^{t,T},$$
and hence, the vega of the gama swap at time $t$ is worth
$$\Upsilon_{\sigma}^{\Gamma}\left(t\right)=2\sigma \E^{2r\tau}\frac{\tau}{T}\frac{S_{t}}{S_{0}}.$$

Concerning the gamma of a gamma swap, we have
$$
\Gamma^{\Gamma} = \frac{2\E^{2rT}}{TS_{0}}\int_{0}^{+\infty}\frac{\D K}{K}\partial_{SS}^{2}O_{K}
= \frac{2\E^{2rT}}{TS_{0}}\int_{0}^{+\infty}\frac{\D K}{K}\frac{\phi\left(d_{1}\right)}{S_{0}\sigma\sqrt{T}}
= 2\frac{\E^{2rT}}{TS_{0}^{2}}
$$
where, for ease of notation, $O_K$ represents either a put or a call of strike $K$.
We here used the same change of variable as for the vega of the gamma swap. 

\section{P\&L of a gamma swap}
We here consider the P\&L of a gamma swap. The calculations below are almost identical to those of Part \textbf{$5.3$}.
$$
\sum_{i=1}^{n}\alpha_{i}P\&L_{i}^{\Gamma}-P\&L_{I}^{\Gamma} = \sum_{i=1}^{n}\frac{1}{2}S_{i}^{2}\left(\frac{\D S_{i}}{S_{i}}-\sigma_{i}^{2}\D t\right)\left(\alpha_{i}\Gamma_{i}-p_{i}^{2}\Gamma_{I}\right)
 +\frac{1}{2}\Gamma_{I}I_{t}^{2}\sum_{i\ne j}w_{i}w_{j}\sigma_{i}\sigma_{j}\left(\rho-\hat{\rho}\right)\D t.
$$
Since the gamma of a gamma swap for a maturity $T$, at time $t$ is $\Gamma=2\exp\left(2r\tau\right)/\left(S_{0}S_{t}T\right)$, we have
\begin{align*}
P\&L_{\text{Disp}}^{\Gamma} & = \frac{1}{2}\sum_{i=1}^{n}\left(\left(\frac{\D S_{i}}{S_{i}}\right)-\sigma_{i}^{2}\D t\right)\left(\alpha_{i}\Gamma_{i}S_{i}^{2}-w_{i}^{2}\Gamma_{I}I_{t}^{2}\right)
 + \frac{1}{2}\Gamma_{I}I_{t}^{2}\sum_{i\ne j}w_{i}w_{j}\sigma_{i}\sigma_{j}\left(\rho-\hat{\rho}\right)\D t\\
 & = \frac{2}{T}\E^{2r\tau}\sum_{i=1}^{n}\left(\left(\frac{\D S_{i}}{S_{i}}\right)-\sigma_{i}^{2}\D t\right)\left(\alpha_{i}\frac{S_{i}}{S_{0}}-w_{i}^{2}\frac{I_{t}}{I_{0}}\right)
 + \frac{1}{T}\E^{2r\tau}\frac{I_{t}}{I_{0}}\sum_{i\ne j}w_{i}w_{j}\sigma_{i}\sigma_{j}\left(\rho-\hat{\rho}\right)\D t.
\end{align*}
The sum that multiplies the correlation spread does not depend on the correlation, but only on the components of the index. Hence, we can define $\beta^{\Gamma}:=T^{-1}\E^{2r\tau}I_0^{-1}I_{t}\sum_{i\ne j}w_{i}w_{j}\sigma_{i}\sigma_{j}$, and eventually
$$P\&L_{\text{Disp}}^{\Gamma}=\frac{2}{T}\E^{2r\tau}\sum_{i=1}^{n}\left(\left(\frac{\D S_{i}}{S_{i}}\right)-\sigma_{i}^{2}\D t\right)\left(\alpha_{i}\frac{S_{i}}{S_{0}}-w_{i}^{2}\frac{I_{t}}{I_{0}}\right)+\beta^{\Gamma}\left(\rho-\hat{\rho}\right)\D t.$$
We can now write the total P\&L of the gamma swap, as well as the one for the dispersion trade via gamma swaps
\begin{align*}
P\&L_{I} & = P\&L_{I}^{\Gamma}+\text{Vega}_{I}\D \sigma_{I}+\frac{1}{2}\text{Volga}_{I}\xi_{I}^{2}\sigma_{I}^{2}\D t+\text{Vanna}_{I}\sigma_{I}I\rho_{I}\xi_{I}\D t\\
 & = P\&L_{I}=P\&L_{I}^{\Gamma}+2\sigma \E^{2r\tau}\frac{I_{t}}{I_{0}}\left(\D \sigma_{I}+\frac{1}{2}\xi_{I}^{2}\sigma_{I}\D t+\rho_{I}\xi_{I}\D t\right).
\end{align*}
Hence
$$
P\&L_{\text{Disp}} = P\&L_{\text{Disp}}^{\Gamma}
 +2\E^{2r\tau}\left(\sum_{i=1}^{n}\alpha_{i}\sigma_{i}\frac{S_{i}}{S_{0}}\left(\D \sigma_{i}+\frac{1}{2}\xi_{i}^{2}\sigma_{i}\D t+\rho_{i}\xi_{i}\D t\right)-\sigma_{I}\frac{I_{t}}{I_{0}}\left(\D \sigma_{I}+\frac{1}{2}\xi_{I}^{2}\sigma_{I}\D t+\rho_{I}\xi_{I}\D t\right)\right).
$$
\section{Arbitrage opportunity condition and vega weighted flat strategy for variance swap dispersion}
We here analyse the vega weighted flat strategy in terms of arbitrage opportunities. We only consider a dispersion trade built upon variance swaps. A portfolio $\alpha=\left(\alpha_{1},\ldots,\alpha_{n}\right)$ represents an arbitrage opportunity if and only if for any $\hat{\sigma}=\left(\hat{\sigma}_{I},\hat{\sigma}_{1},\ldots,\hat{\sigma}_{n}\right)$, we have
$$\sigma_{I}^{2}-\hat{\sigma}_{I}^{2}+\sum_{i=1}^{n}\left(\hat{\sigma}_{i}^{2}-\sigma_{i}^{2}\right)\geq 0.$$
Rearranging the terms, we have
$$\left(\sum_{i=1}^{n}\hat{\sigma}_{i}^{2}-\hat{\sigma}_{I}^{2}\right)+\left(\sigma_{I}^{2}-\sum_{i=1}^{n}\sigma_{i}^{2}\right)\geq 0.$$
In particular, if $\hat{\sigma}=0$, then $\sigma_{I}^{2}-\sum\limits_{i=1}^{n}\alpha_{i}\sigma_{i}^{2}\geq 0$, i.e.,
$\left(\sum_{i=1}^{n}\alpha_{i}\sigma_{i}^{2}\right)/\sigma_{I}^{2}\leq 1$.
If we consider the vega weighted flat strategy, we have 
$\alpha_{i}=w_{i}\sigma_{I}^{2}/\left(\sigma_{i}\sum_{j=1}^{n}\alpha_{j}\sigma_{j}\right)$, and with this weighting schemes, we see that
$\left(\sum_{i=1}^{n}\alpha_{i}\sigma_{i}^{2}\right)/\sigma_{I}^{2}=1$, and the vega weighted flat strategy represents the boundary condition for arbitrage opportunity. 
\end{document}